\documentclass[11pt]{article}

\usepackage[preprint]{acl}

\usepackage{times}
\usepackage{latexsym}

\usepackage[T1]{fontenc}

\usepackage[utf8]{inputenc}

\usepackage{microtype}

\usepackage{inconsolata}

\usepackage{graphicx}
\usepackage{booktabs}
\usepackage{xspace}
\usepackage{xcolor}
\usepackage{colortbl}
\usepackage[frozencache,cachedir=.]{minted}
\usepackage{minted}
\usepackage{amsmath}
\usepackage{mathpartir}
\usepackage{algorithm}
\usepackage{algpseudocode}
\usepackage{subcaption}
\usepackage{enumitem}

\newcommand{\sysname}{BinSeek\xspace}
\newcommand{\sysemb}{BinSeek-Embedding\xspace}
\newcommand{\sysrank}{BinSeek-Reranker\xspace}
\newcommand{\ie}{\textit{i.e.,}\xspace}
\newcommand{\eg}{\textit{e.g.,}\xspace}

\title{Cross-modal Retrieval Models for Stripped Binary Analysis}

\author{
 \textbf{Guoqiang Chen\textsuperscript{1,2}},
 \textbf{Lingyun Ying\textsuperscript{2}},
 \textbf{Ziyang Song\textsuperscript{2}},
 \textbf{Daguang Liu\textsuperscript{2}},
\\
 \textbf{Qiang Wang\textsuperscript{2}},
 \textbf{Zhiqi Wang\textsuperscript{2}},
 \textbf{Li Hu\textsuperscript{1}},
 \textbf{Shaoyin Cheng\textsuperscript{1}},
\\
 \textbf{Weiming Zhang\textsuperscript{1}},
 \textbf{Nenghai Yu\textsuperscript{1}}
\\
 \textsuperscript{1}University of Science and Technology of China,
 \textsuperscript{2}QI-ANXIN Technology Research Institute
}

\begin{document}
\maketitle
\begin{abstract}
Retrieving binary code via natural language queries is a pivotal capability for downstream tasks in the software security domain, such as vulnerability detection and malware analysis.
However, it is challenging to identify binary functions semantically relevant to the user query from thousands of candidates, as the absence of symbolic information distinguishes this task from source code retrieval.
In this paper, we introduce, \sysname, a two-stage cross-modal retrieval framework for stripped binary code analysis. It consists of two models: \sysemb is trained on large-scale dataset to learn the semantic relevance of the binary code and the natural language description, furthermore, \sysrank learns to carefully judge the relevance of the candidate code to the description with context augmentation. 
To this end, we built an LLM-based data synthesis pipeline to automate training construction, also deriving a domain benchmark for future research. 
Our evaluation results show that \sysname achieved the state-of-the-art performance, surpassing the the same scale models by 31.42\% in Rec@3 and 27.17\% in MRR@3, as well as leading the advanced general-purpose models that have 16 times larger parameters. 
\end{abstract}

\section{Introduction}

Binary code analysis serves as a cornerstone of software security, supporting critical tasks such as vulnerability detection, malware analysis, program audit. 
Traditional manual analysis is known to be labor-intensive and requires deep expertise, especially for binaries including thousands of functions. 
To alleviate these scalability challenges, recent studies~\cite{BinMetric, shang2024bcodeunderstanding} have begun exploring the application of Large Language Models (LLMs) to various binary code-related analysis tasks. Nevertheless, existing work has largely overlooked the problem of retrieving binary code through natural language (NL) queries, leaving a gap in effectively bridging users' high-level semantic intentions with large-scale binary codebases.

Retrieving stripped binary code with NL queries, however, presents unique and substantial challenges compared to source code retrieval.
Unlike high-level programming language (PL), stripped binaries lack explicit semantic indicators such as variable names, function names, and type information.
These comprehensive symbols are generally stripped before releasing the binary programs for different reasons (\eg reducing file size, hiding functionality).
This absence significantly complicates the understanding of binary code for both humans and models~\cite{BinSum-jin2023}, downgrading the performance of general-purpose retrieval models in this scenario. 

Semantic-based code search has been extensively explored in the context of source code. 
Existing approaches typically learn a joint embedding space to align NL and code, with pretrained models such as CodeBERT~\cite{feng2020CodeBERT}, GraphCodeBERT~\cite{guo2021GraphCodeBERT}, and UniXcoder~\cite{guo2022unixcoder} achieving remarkable success. 
Despite these advancements, directly adapting source code retrieval models to the binary domain is ineffective due to the fundamental modality differences discussed above.
Moreover, unlike the abundance of open-source repositories (\eg GitHub) that fuel source code representation learning, a high-quality and well-labeled dataset for binaries remain scarce, constraining the development of robust binary code retrieval models. 

Since retrieving binary code with NL queries is rarely studied, this work aims to develop an effective solution for retrieving functions from stripped binaries. 
In particular, we introduce \sysname, a two-stage cross-modal retrieval framework specifically tailored for this issue: for the first stage, we build, \sysemb, a embedding model to retrieve the candidates from a codebase that is decompiled from binaries, and for the second stage, we further devise, \sysrank, to reorder the candidates augmented with calling context information for more precise results. 
To train our expert models, we employ LLMs to automatically synthesize high-quality semantic labels in NL for binary functions. 
In this way, we also deliver the first domain benchmark for the binary code retrieval task, which is expected to facilitate future research in this domain. 
Our \sysname achieves state-of-the-art (SOTA) performance with Rec@3 of 84.5\% and MRR@3 of 80.25\%, indicating its effectiveness in finding semantic-related binary functions. 
Our main contributions are summarized as follows: 
\begin{itemize}[left=0.2cm] 
    \setlength{\itemsep}{0pt}
    \vspace{-1.4ex}
    \item We introduce \sysname, a two-stage cross-modal retrieval framework for stripped binary code analysis, where \sysemb is used to retrieve candidates from a codebase, and \sysrank further refine candidates with calling context information.
    \vspace{-0.5ex}  
    \item We propose an LLM-driven data synthesis pipeline to automate the construction of high-quality training data, also deriving the first domain benchmark for binary code retrieval.
    \vspace{-0.5ex}
    \item Our evaluations demonstrate that \sysname achieved SOTA performance, significantly outperforming baselines with superior efficiency. 
\end{itemize}
\vspace{-1.0ex}
We make our model and evaluation benchmark publicly available at \href{https://github.com/XingTuLab/BinSeek}{here}.

\section{Background and Challenges}
\vspace{-2.0ex}
\label{sec:background}

\begin{figure}[t]
\centering
\captionsetup{skip=1ex}
\begin{subfigure}{\linewidth}
\centering
\begin{minted}[frame=lines,linenos,numbersep=5pt,breaklines=true,fontsize=\small,escapeinside=||]{c}
static void xtea_encode(uint32_t num_rounds, 
  uint32_t ori[2], uint32_t const key[4]){
  unsigned int i;
  uint32_t data0 = ori[0], data1 = ori[1],     
    sum = 0, delta = 0x9E3779B9;
  for (i = 0; i < num_rounds; i++){
    data0 += ...;
    sum += delta;
    data1 += ...; }
  ori[0] = data0; ori[1] = data1;
}
\end{minted}
\vspace{-0.4cm}
\caption{Source code function for XTEA cryptographic algorithm.}
\vspace{0.2cm}
\end{subfigure}

\begin{subfigure}{\linewidth}
\centering
\begin{minted}[frame=lines,linenos,numbersep=5pt,breaklines=true,fontsize=\small,escapeinside=||]{c}
__int64 __fastcall sub_100000A68(__int64 result, 
  unsigned int *a2, __int64 a3){
  unsigned int v3; // [xsp+8h] [xbp-28h]
  unsigned int v4; // [xsp+Ch] [xbp-24h]
  unsigned int v5; // [xsp+10h] [xbp-20h]
  unsigned int i; // [xsp+14h] [xbp-1Ch]
  v5 = *a2; v4 = a2[1]; v3 = 0;
  for (i = 0; i < (unsigned int)result; ++i){
    v5 += ...;
    v3 -= 1640531527;
    v4 += ...; }
  *a2 = v5; a2[1] = v4;
  return result;
}
\end{minted}
\vspace{-0.4cm}
\caption{Binary pseudocode decompiled by IDA Pro.}
\end{subfigure}

\caption{
Example of source code and corresponding decompiled pseudocode. 
}
\vspace{-3.5ex}
\label{fig:example-code}
\end{figure}

\subsection{Problem Definition}

In this work, we investigate the problem of retrieving binary functions using NL queries. Let $\mathcal{P}$ denote the set of pseudocode decompiled from stripped binary files, where each $p_i \in \mathcal{P}$ corresponds to a binary function in the codebase. Given an NL query $q$ describing a desired behavior or functionality, the objective is to search the most semantically relevant pseudocode functions:
\vspace{-1.5ex}
\begin{equation}
\mathcal{R}(q, \mathcal{P}) = \{p_{1}, p_{2}, \ldots, p_{k}\}
\end{equation}
\vspace{-4ex}

\noindent This formulation reflects practical scenarios, where human or LLM-based analysts aim to effectively and quickly locate relevant functionality within stripped binaries using high-level descriptions and insights, for tasks such as vulnerability investigation, malware analysis, and software auditing.
We focus on decompiled binary code rather than disassembly code, as previous studies have highlighted it is more human-readable and LLM-friendly representation~\cite{NER-2023pst, BinSum-jin2023}. 

This task is distinct from binary code similarity detection (BCSD). BCSD methods (\eg jTrans~\cite{Wang2022jTrans} and CLAP~\cite{Wang2024CLAP}) focuses on measuring semantic similarity between two binary code snippets, typically to determine whether they implement identical or related functionality. In contrast, our task is inherently cross-modal: the input is NL, while the target consists of pseudocode functions. The core challenge lies in bridging the semantic gap between high-level linguistic intent and the semantics implicitly preserved in stripped binary code. 

\begin{figure*}[t]
    \centering
    \captionsetup{skip=2pt}
    \scalebox{0.92}{
    \includegraphics[width=1\linewidth]{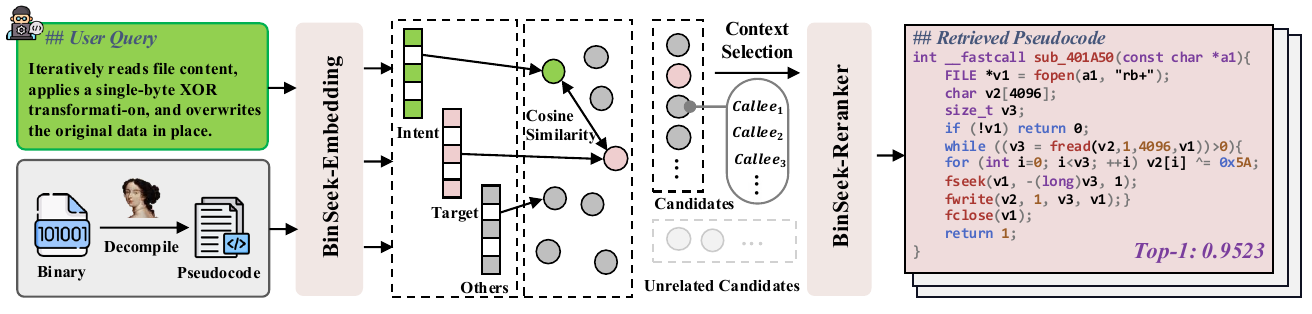}
    }
    \caption{Overview of \sysname.}
    \vspace{-3.0ex}
    \label{fig:overview}
\end{figure*}

\subsection{Source Code Retrieval Model}

Code retrieval models aim to learn a shared embedding space where both NL descriptions and PL code are mapped to dense vector representations, enabling similarity-based retrieval of relevant code snippets. 
CodeBERT~\cite{feng2020CodeBERT} presented the first bimodal pre-trained model for PL and NL, GraphCodeBERT~\cite{guo2021GraphCodeBERT} further introduced data flow information to enhance code representation learning. UniXcoder~\cite{guo2022unixcoder} leverages cross-modal contents like abstract syntax tree and code comment for more improvement, while CodeXEmbed~\cite{liu2024codexembed} designed a family of code embedding models tailored to diverse PLs and retrieval tasks. 
Furthermore, many comprehensive benchmarks and datasets like CodeSearchNet~\cite{husain2019codesearchnet}, CodeXGLUE~\cite{Lu2021CodeXGLUE}, and CoIR~\cite{CoIR} have facilitated the evaluation of these models, demonstrating their strong capabilities in searching from a large-scale corpus. 
However, similar research in the binary code domain remains limited, which motivates our current investigation into cross-modal retrieval models for binary analysis. 

\vspace{-1ex}
\subsection{Stripped Binary Code Retrieval}
\label{sec:background-bcode}
\vspace{-0.5ex}

Binary code is a low-level representation of a program, typically compiled from the source code written in PLs like C/C++, and it consists of a sequence of machine instructions encoded in binary format. Due to various reasons (\eg reducing file size, hiding functionality), most binary files are stripped of symbol information before release, resulting in the lacks of function names, variable names, data types, and other high-level semantic information. Although modern decompilers can handle binaries and generate C-like decompiled pseudocode, they use placeholders to represent missing symbolic information, like the function \texttt{sub\_100000A68} and the variable \texttt{a2/v3} shown in \autoref{fig:example-code}.

Recently, many studies have explored automated analysis of binary pseudocode using LLMs, such as LLM4Decompile~\cite{tan2024llm4decompile}, ReCopilot~\cite{chen2025recopilot}, and Bin2Wrong~\cite{usenix2025Bin2Wrong}, yielding promising experimental results. 
In practice, however, a binary program usually contains tens of thousands of functions~\cite{Cisco2022HowMachine, Sun2023GraphMoCo}, highlighting the difficulty of searching for specific code within it. 
To our knowledge, the only related work, BinQuery~\cite{binquery}, conducted investigation on retrieving disassembled code with NL queries (distinct from our problem definition). They demonstrated the potential of learning-based methods, though the performance remain unsatisfactory. 
It is particularly difficult given the absence of high-level semantic symbols in binary code. 
Specifically, we summarize the critical challenges as follows:
\begin{itemize}[left=0.2cm] 
    \setlength{\itemsep}{0pt}
    \vspace{-1.4ex}
    \item \textbf{Lacking of Symbols:} Stripping symbols creates a significant gap between the decompiled pseudocode and the source code, making it harder for the model to identify the semantics. Consequently, this gap prevents us from reusing the source code retrieval models. 
    \vspace{-0.5ex}
    \item \textbf{Irreversible Semantics Loss:} Given the information loss during compilation and stripping, it is typically hard to infer the source-level semantics of a function solely from its binary code. For instance, the function name in the source code in \autoref{fig:example-code} indicates its operational context for cryptographic encoding, which cannot be deduced from the binary function alone. Therefore, effective binary code retrieval models must learn to infer semantics from the context. 
\end{itemize}

\vspace{-2ex}
\section{Method}
\vspace{-1ex}


\autoref{fig:overview} illustrates the overall pipeline of \sysname, which comprises the \sysemb and \sysrank models. 
\sysemb maps user queries in NL and pseudocode functions into embeddings, measuring cross-modal semantic relevance based on the embedding distances. 
Each vector requires only one single model computation, making \sysemb suitable for fast retrieval over large codebases. 
However, as discussed in \S\ref{sec:background-bcode}, \sysemb alone may struggle to capture complete semantics from a pseudocode alone, resulting in suboptimal retrieval performance. 
To mitigate this, we introduce \sysrank as a secondary reranking step. 
\sysrank leverages the supplementary information selected from the calling context for a more comprehensive inference of function semantics, thereby optimizing the retrieval results.

\subsection{Data Synthesis}
\label{sec:data-synthesis}

Training cross-modal retrieval models requires a large amount of data pairs that consist of decompiled pseudocode and corresponding semantic descriptions in NL. Unfortunately, there is currently no publicly available dataset specifically designed for binary code retrieval. 
Although hiring reverse engineering experts to label pseudocode is the optimal approach for the best quality, high labor costs make it impractical for large-scale execution. 
To overcome it, we propose a data synthesis pipeline to automatically generate high-quality training data. 

\begin{figure}[htbp]
    \vspace{-1ex}
    \centering
    \scalebox{0.9}{
    \includegraphics[width=1\linewidth]{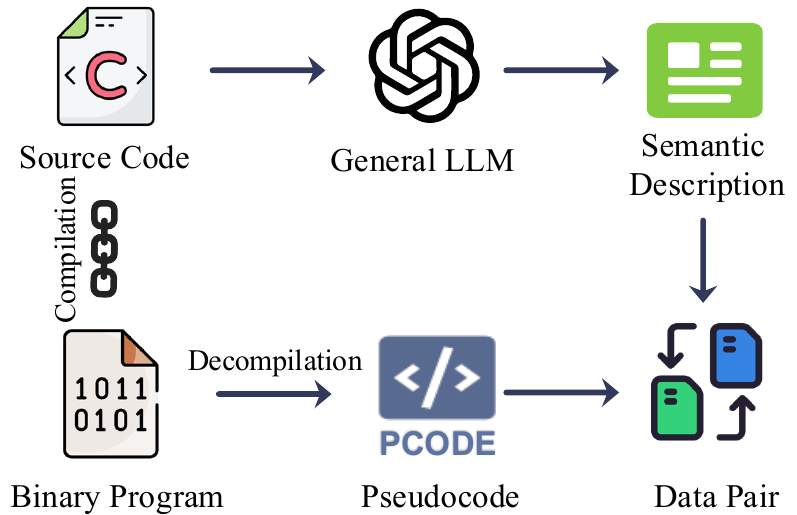}
    }
    \vspace{-1.0ex}
    \caption{Pipeline of data synthesis.}
    \vspace{-2.0ex}
    \label{fig:data-synthesis}
\end{figure}


As shown in \autoref{fig:data-synthesis}, we first collect a large number of open-source projects written in C/C++ from GitHub. For each project, we compile the source code into stripped binary files using different configurations, while saving the debugging information separately (more details in \S\ref{sec:appendix-raw-data}). 
Next, we utilize a modern decompiler (\ie IDA Pro~\cite{IDA}) to decompile binary files into pseudocode functions. Since stripped binaries do not have symbols, the debugging information serves as a bridge linking the source code to the binary code in the function level. 
Finally, we employ an advanced LLM (\ie DeepSeek~\cite{2025deepseekv3}) to generate semantic descriptions for each source code function. 
More specifically, we provide the LLM with the content of the source code file containing the function, along with the file path, project name, and version number as references (detailed in \S\ref{sec:appendix-data-syn}). 
NL descriptions, source code, and binary code are essentially representations of the same semantics across different modalities.
We therefore synthesize the training data pairs with generated NL descriptions and binary code.

\noindent\textbf{Quality Filters.} To ensure the quality of the synthesized data, we eliminate low-quality data pairs with the following filters:
\vspace{-1ex}
\begin{itemize}[left=0.2cm] 
    \setlength{\itemsep}{0pt}
    \vspace{-1.0ex}
    \item \textbf{Source Code Filter} drops the function with too short lines of code (LoC$\leq$10), indicating lack of meaningful semantics. 
    \vspace{-1.5ex}
    \item \textbf{Binary Code Filter} also removes short binary functions similarly, as well as thunk functions and virtual functions, which typically is the wrapper for calling other functions.
    \vspace{-1.0ex}
    \item \textbf{Generation Filter} prompts an LLM to assess the quality of the generated NL description, filtering out those with errors, irrelevant content, or redundant information (see \S\ref{sec:appendix-data-syn}). 
    \vspace{-1.0ex}
    \item \textbf{Repetitive Semantics Filter} drops both source and binary functions with high similarity to reduce code redundancy. We employ MinHash~\cite{minhash}, a locality-sensitive hashing algorithm, to perform function-level deduplication, and use the empirical threshold of 0.95~\cite{TOSEM2025FoC}.
\end{itemize}
\vspace{-1ex}
\noindent Ultimately, we constructed \texttt{106,711,414} raw data pairs in total. 
A human evaluation shows that 97.6\% of the generated descriptions are rated as correct or at least partially correct, indicating the reliability of the synthesized labels (detailed in \S\ref{sec:human-eval}). 

\noindent\textbf{Data Sampler.} 
Training retrieval models typically require both positive and negative data pairs. 
The positive pair naturally consist of the NL description and pseudocode function from the same source code. 
To construct effective negative samples, we employ a hybrid strategy combining random sampling and hard negative mining~\cite{hardnegativemining2025CIKM}. 
Specifically, for each source code function, we take the NL description as a query and randomly select a bunch of binary functions as negative candidates that are compiled from the other source code among the whole corpus. 
Furthermore, we utilize a general embedding model (\texttt{Qwen3-Embedding-8B}~\cite{2025qwen3-emb}), which is top-tier in the MTEB leaderboard~\cite{2025mmteb}, to estimate the semantic similarity between the query and the NL descriptions of the negative candidates. 
After arranging in descending order of semantic similarity, we select the first binary function with a score below the threshold ($\leq 0.95$) as a hard negative sample. 
In this way, we constructed \texttt{45,703,009} high-quality positive and negative data pairs in total, enabling the model to learn the mappings across different semantic representations.

\vspace{-0.5ex}
\subsection{Model Architecture}
\vspace{-0.5ex}

\noindent\textbf{\sysemb.} The embedding model is built from scratch with reference to the existing language model Qwen3~\cite{qwen3}. 
Specifically, we initialize our model backbone with an 8 transformer layer, which incorporates RMSNorm~\cite{RMSNorm}, SwiGLU activation~\cite{SwiGLU}, and grouped query attention (GQA)~\cite{GQA} techniques for improved training stability and performance. 
We employ Qwen3's tokenizer that implements byte-level byte-pair encoding~\cite{Wang_Cho_Gu_2020, sennrich-etal-2016-neural} with a vocabulary size of 151,669. 
Consequently, the weights of the token embedding layer are directly loaded from the well-trained Qwen3 model and frozen to reduce training overhead. 
Following the embedding layer, we use a rotary position embedding (RoPE)~\cite{RoPE} layer to provide position information to transformer blocks, meanwhile enabling input length scaling at inference time.
Finally, the last token pooling is employed to the backbone output to generate the final embedding. 


\noindent\textbf{\sysrank.} The reranking model is built upon the \sysemb with 18 transformer layers, further stacking one Language Modeling (LM) head on the top. 
The LM head outputs the probability logits for the input code and NL query sequence, which are then transformed by a sigmoid function to produce a score ranging from 0 to 1, indicating the likelihood of semantic relevance.
In practice, \sysrank receives the user query and retrieves candidates from \sysemb, and it boosts the overall accuracy of the system through reranking. 


\noindent\textbf{Context-aware Design.} 
Since the pseudocode of a single binary function can only provide limited information (mentioned in \S\ref{sec:background-bcode}), we make the contextual functions available to the model in the reranking stage for extending its semantic inference capabilities. 
As shown in \autoref{fig:overview}, we assemble the target function $f_t$ and its callees $f_c$ into the whole input. However, the callees could largely increase the input length, even pose the out-of-context (OOC) problem, leading to inefficiency and potential noise. 

To mitigate this, we here propose a heuristic function importance metric to select the top-$k$ important callees to provide context. 
Existing studies have demonstrated that the presence of meaningful function names and strings often provides significant semantic cues about its functionality~\cite{NER-2023pst, BinMetric, BinSum-jin2023}. Therefore, we try to measure the meaningful symbol density of a pseudocode function $f_t$ through three aspects: (1) whether the function has an unstripped function name, (2) the density of strings among the pseudocode, and (3) the meaningful callee names in the function. 
Formally, we define the informative score $\mathcal{S}(f_t)$ to indicate function importance as follows:
\vspace{-1ex}
\begin{equation}
    \footnotesize
    \mathcal{N}(f) = 
    \begin{cases} 
        1, & \text{exists name symbol of } f \\
        0, & \text{otherwise}
    \end{cases}
\end{equation}
\vspace{-2ex}
\begin{equation}
    \footnotesize
    \mathcal{S}(f_t) = \mathcal{N}(f_t) + 
    \sigma \left( \beta \cdot \frac{\text{count}(T_{str})}{\text{count}(T_{code})} \right) + 
    \frac{1}{n} \sum_{i=1}^{n} \mathcal{N}(f_c^i) 
\end{equation}
\noindent where $\mathcal{N}(f)$ is an indicator of the existence of a function name and $\beta$ is a scaling factor for string density measurement. We empirically set $\beta$ to 15, means the pseudocode with 7.3\% strings tokens will contribute 0.5 string score. And $f_c^i$ is the $i$-th callee function among the $n$ callees in the function $f_t$. In general, $\mathcal{S}(f_t)$ ranges from 0 to 3, with higher values indicating more informative functions. 
Finally, we set $k=5$ to selectively preserve callee pseudocode as context augmentation. 

\subsection{Model Training}

In this section, we describe the training details of the \sysemb and \sysrank models.
We first present the loss functions used for model optimization, followed by specific training configurations and hyperparameters. 

\noindent\textbf{Loss Function.}
We employ different loss functions for the two models due to their distinct architectures and training objectives, as shown in \autoref{fig:overview}. For the \sysemb model, we utilize InfoNCE loss~\cite{2019InfoNCE} to optimize the cross-modal embedding space. Given a batch of $N$ positive pairs of NL descriptions and binary functions, we compute their embeddings using the \sysemb model. 
The InfoNCE loss is defined as follows: 
\vspace{-1ex}
\begin{equation}
    \footnotesize
    \mathcal{L}_{e} = - \frac{1}{N} \sum_{i=1}^{N} \log \frac{\exp(\text{sim}(q_i, p_i) / \tau)}{\sum_{p_j \in \{p_j^-\} \cup \mathcal{P}} \exp(\text{sim}(q_i, p_j) / \tau)}
\end{equation}
\noindent where $q_i$ and $p_i$ are the embeddings of the NL description and pseudocode function in the $i$-th positive pair, respectively, while the $p_j^-$ is the hard negative. The $\text{sim}(\cdot, \cdot)$ denotes the similarity function, which is cosine similarity in our practice, and $\tau$ is a temperature hyperparameter controlling the concentration level of the distribution.

As for \sysrank model, we adopt the binary cross-entropy (BCE) loss to optimize the relevance prediction between NL descriptions and context-integrated binary functions. Given a batch of $M$ samples consisting of both positive and negative pairs, the BCE loss is defined as follows:
\vspace{-1ex}
\begin{equation}
    \footnotesize
    \mathcal{L}_{r} = - \frac{1}{M} \sum_{i=1}^{M} \left[ y_i \log(p_i) + (1 - y_i) \log(1 - p_i) \right]
\end{equation}
\noindent where $y_i$ is the ground-truth label indicating whether the $i$-th pair is relevant ($1$) or not ($0$), and $p_i$ is the predicted relevance score output by \sysrank model.

\noindent\textbf{Training Setup.}
We train both models using the AdamW optimizer with a learning rate of \texttt{1e-4} and a cosine weight decay scheduler with minimum of \texttt{1e-5}.
For the \sysemb, we use an equivalent batch size of \texttt{5,120} and train for \texttt{2} epochs, with a maximum sequence length of \texttt{4,096} tokens. 
The \sysrank is trained with an equivalent batch size of \texttt{512} for \texttt{3} epochs, enlarging a maximum sequence length to \texttt{16,384} tokens for holding contextual information. 
Our experiments are conducted on H100 $\times$ 8 GPUs, each equipped with 80GB GPU memory, 2TB RAM, and \texttt{Intel Xeon Platium 8468} with \texttt{48} cores.
The whole training of \sysname requires approximately \texttt{2k} GPU hours in total. 

\vspace{-1ex}
\section{Benchmarking Binary Code Retrieval}
\vspace{-1ex}


Since \sysname incorporated two distinct models, we build a comprehensive benchmark dataset to assess retrieval and reranking performance separately. 
We split 10\% of the raw dataset for test data construction, and ensure that the source code of these data does not overlap with training data to prevent data leakage. 
In total, we constructed 400 test data for embedding evaluation and 400 for reranking evaluation, and the details are as follows:

\noindent\textbf{Evaluate Embedding Model.}
To evaluate the \sysemb's retrieval capability, we created a search pool containing large-scale binary functions for each NL query, mirroring its intended real-world application. 
Specifically, for each positive pair $(q_i, p_i)$ in the synthesized corpus, where $q_i$ denotes the NL description and $p_i$ denotes the corresponding pseudocode function, we constructed a search pool $\mathcal{P}_i$ of size $|\mathcal{P}_i| = K$ (empirically set to 10,000 in our experiments). 
The pool comprises the positive sample $p_i$ along with $(K-1)$ carefully selected negative samples that are semantically unrelated to the positive one. 

Given a candidate $p_j$ from the corpus, we assess its semantic relevance to the positive sample $p_i$ by comparing their associated NL descriptions. 
Following our training data sampler (\S\ref{sec:data-synthesis}), we employed the same threshold of similarity score for filtering. This design ensures that no unlabeled positive sample leaks into the negative samples. 
Notably, we further employ an LLM to generate rephrased semantic descriptions for the candidates, thereby obtaining a more comprehensive measure of correlation. 
Algorithm \autoref{alg:embedding_eval} of \S\ref{sec:appendix-data-const-algo} comprehensively presents the processing for iteratively collecting test data with similarity-based filtering.




\noindent\textbf{Evaluate Reranking Model.}
Our \sysrank model is designed to rerank the candidates retrieved by an embedding model. To assess the reranking performance, we construct the derived evaluation dataset upon the one built above. Specifically, for each test data $(q_i, \mathcal{P}_i)$, we first employ our trained \sysemb model to retrieve the top-N candidates from the pool $\mathcal{P}_i$ (empirically set $N=10$ in our experiments). If only the positive sample $p_i$ is retrieved within the candidates but not the top-1 candidate, we take the positive sample and $(N-1)$ negative samples as a reranking test data $(q_i, \mathcal{R}_i)$. This evaluation dataset can effectively measure the capability of rerankers in refining retrieval results and improving overall accuracy. Algorithm \autoref{alg:reranking_eval} of \S\ref{sec:appendix-data-const-algo} outlines the complete procedure for constructing the reranking model evaluation dataset.



In summary, we sampled a moderately sized test dataset from a large-scale test corpus, with the primary goal of ensuring practical usability and ease of adoption. To verify its representativeness, we reconstructed the test set via repeated resampling from the raw corpus, observing a standard deviation of 1.52\% in our evaluation metric across trials.
Furthermore, repeated subsampling of the fixed final test queries yields a lower standard deviation of 0.91\%, indicating stable evaluation results.
Together, these results suggest that our test dataset achieves a favorable balance between usability, representativeness, and statistical stability.

\begin{table*}[htbp]
\centering
\small
\setlength{\tabcolsep}{3pt}
\caption{The evaluation results of embedding models.}
\vspace{-2.0ex}
\rowcolors{2}{gray!15}{white}
\begin{tabular}{@{}llccccc@{}}
\toprule
Model                          & Size (B) & \multicolumn{1}{l}{$Rec@1$} & \multicolumn{1}{l}{$Rec@3$} & \multicolumn{1}{l}{$Rec@10$} & \multicolumn{1}{l}{$M\!R\!R@3$} & \multicolumn{1}{l}{$M\!R\!R@10$} \\ \midrule
embeddinggemma-300m            & 0.3      & 40.50                        & 51.00                        & 58.50                         & 45.25                     & 46.65                      \\
SFR-Embedding-Code-400M\_R     & 0.4      & 22.00                        & 29.50                        & 35.00                         & 25.42                     & 26.41                      \\
BGE-M3                         & 0.6      & 26.50                        & 32.50                        & 36.00                         & 29.17                     & 29.90                      \\
Qwen3-Embedding-0.6B           & 0.6      & 51.00                        & 59.00                        & 66.00                         & 54.50                     & 55.71                      \\
multilingual-e5-large-instruct & 0.6      & 32.50                        & 40.50                        & 46.00                         & 36.00                     & 36.82                      \\
jina-embeddings-v4             & 4        & 41.50                        & 49.50                        & 59.50                         & 45.42                     & 47.24                      \\
inf-retriever-v1               & 7        & 51.50                        & 59.00                        & 67.50                         & 54.58                     & 56.07                      \\
SFR-Embedding-Mistral          & 7        & \underline{60.50}            & \underline{69.50}            & \underline{77.50}             & \underline{64.67}         & \underline{66.16}          \\
Qwen3-Embedding-8B             & 8        & 57.50                        & 65.00                        & 73.50                         & 60.75                     & 62.14                      \\
\sysemb                        & 0.3      & \textbf{67.00}               & \textbf{80.50}               & \textbf{93.50}                & \textbf{72.83}            & \textbf{75.18}             \\ \bottomrule
\end{tabular}
\label{tab:emb-compare}
\vspace{-4ex}
\end{table*}

\vspace{-1ex}
\section{Experiments}
\label{sec:exp}
\vspace{-1ex}

We conducted extensive experiments to evaluate the effectiveness of our proposed \sysname. We here to answer the following research questions: 
\begin{itemize}[left=0.2cm] 
    \setlength{\itemsep}{0pt}
    \vspace{-1.5ex}
    \item \textbf{RQ1:} How do our \sysemb and \sysrank perform compared to other models in the cross-modal retrieval of stripped binary functions?
    \vspace{-0.5ex}
    \item \textbf{RQ2:} How does our \sysname perform as a system in practice?
\end{itemize}

\vspace{-2.5ex}
\subsection{Evaluation Settings}
\label{sec:eval-settings}
\vspace{-0.8ex}

\noindent\textbf{Metrics.} 
We employ two widely-adopted metrics to comprehensively measure the embedding and reranking models. 
Recall at $k$ (Rec@k) quantifies the proportion of queries for which the relevant item appears within the top-$k$ retrieved results, directly indicating the system's success rate in retrieving the correct function among the top candidates.
While, the Mean Reciprocal Rank at $k$ (MRR@k) further focuses on the rank position of the postive item, considering only the top-$k$ results for each query. MRR@k computes the average of the reciprocal ranks of the positive item, providing insight into how highly the correct function is ranked within the top candidates.

We report the metrics at $k = \{1, 3, 10\}$, which are chosen to reflect different practical scenarios in our retrieval task. 
Specifically, $k=1$ corresponds to the strictest setting, as each query is associated with a single positive instance; this setting evaluates the model's precision in ranking the correct result at the very top. 
The choice of $k=3$ is motivated by practical tolerance for retrieval errors, where returning the top three results is considered acceptable. 
Finally, considering the whole retrieval-reranking system, we set $k=10$ to keep the initially retrieved candidates, who can be passed to the reranking model for further refinement.


\noindent\textbf{Baselines.} 
To the best of our knowledge, the only related work is BinQuery~\cite{binquery}, which, however, does not publicly release their model weights and dataset, making direct comparison infeasible. Moreover, BinQuery focuses on disassembly code, which differs from our NL-pseudocode retrieval task. Therefore, there are no publicly available domain-specific methods that can be directly compared with ours. 
As a result, we conduct a comparison against existing general-purpose models. In selecting baselines, we consider factors including the SOTA performance in public leaderboards~\cite{2022mteb, 2025mmteb, CoIR}, popularity among community, and accessibility for reproduction. 
Finally, we selected a total of 13 baseline models: 9 embedding models and 4 reranking models, detailed in \autoref{tab:baselines} of \S\ref{sec:appendix-model-select}. 





\vspace{-1ex}
\subsection{Embedding \& Reranking Evaluation}
\vspace{-0.5ex}

To answer \textbf{RQ1}, we conducted a comprehensive evaluation of our retrieval framework. We first assess the standalone retrieval performance of \sysemb and the baselines. The test dataset has been outlined in \S\ref{sec:data-synthesis}, with each query containing one positive and 9,999 negatives.

\noindent\textbf{Embedding Models.} 
As shown in \autoref{tab:emb-compare}, \sysemb achieves the best performance across all metrics, establishing a new SOTA for stripped binary code retrieval. Compared to the strongest baseline SFR-Embedding-Mistral (7B parameters), our model (0.3B parameters) achieves significant improvements: 
6.5\% higher in Rec@1 and 16.0\% higher in Rec@10. The MRR@10 is also improved by 9.02\%, demonstrating superior ranking quality of the retrieved results. 

First, we analyze the performance of \sysemb on different metrics to understand its retrieval capability. 
The Rec@1 score of 67\% demonstrates essentially strong performance in large-scale retrieval. 
The Rec@3 score indicates that our model can directly return useful results with an 80.5\% probability in practical acceptance. 
More importantly, Rec@10 achieves 93.5\%, suggesting that the vast majority of suboptimal results have the opportunity to be corrected to top-3 or even top-1 through subsequent reranking.
The MRR@10 of 75.18\% further confirms that the correct results are typically ranked at high positions, demonstrating the semantic relevance of retrieved results.

\begin{table*}[t]
    \captionsetup{skip=2pt}
    \centering
    \small
    \caption{The overall evaluation results of \sysname with different embedding and reranking models.}
    \setlength{\belowcaptionskip}{2pt}
    \setlength{\tabcolsep}{2pt}
    \begin{tabular}{@{}llll|cccccl@{}}
        \toprule
        \multicolumn{2}{c}{Embedding Stage}                 & \multicolumn{2}{c|}{Reranking Stage}  & $Rec@1$             & $Rec@3$          & $Rec@10$          & $M\!R\!R@3$              & $M\!R\!R@10$           & Time(m) \\ \midrule
        \multicolumn{1}{c}{Model} & \multicolumn{1}{c}{Sz.(B)} & \multicolumn{1}{c}{Model} & \multicolumn{1}{c|}{Sz.(B)} &          &          &           &       &        &          \\ \midrule
        embeddinggemma-300m       & 0.3                     & Qwen3-Reranker-0.6B       & 0.6        & 49.00               & 53.08              & 58.75             & 53.08              & 53.34            & \underline{1.91}    \\
        SFR-Embedding-Mistral     & 7                       & Qwen3-Reranker-8B         & 8          & \underline{73.75}   & \underline{77.75} & \underline{78.25}  & \underline{75.62}  & \underline{75.73}          & 20.93    \\
        BinSeek-Embedding         & 0.3                     & BinSeek-Reranker          & 0.6        & \textbf{76.75}      & \textbf{84.50}     & \textbf{93.00}    & \textbf{80.25}     & \textbf{81.51}  & \textbf{1.76}    \\ \bottomrule
        \end{tabular}
    \label{tab:pipeline-eval}
\vspace{-4ex}
\end{table*}

Second, we compare \sysemb with baselines to highlight its advantages in both accuracy and efficiency.
We first examine models with similar parameter scale. Despite having a close size (0.3B), \sysemb significantly outperforms embeddinggemma-300m, achieving 26.5\% higher in Rec\!@\!1 and 28.53\% higher in MRR\!@\!10. 
More remarkably, compared to the best baseline \texttt{SFR-Embedding-Mistral} that has $23\times$ more parameters, \sysemb still achieves 16\% higher Rec\!@\!10 and 9\% higher MRR\!@\!10. 

\noindent\textbf{Reranking Models.} 
In this evaluation, each reranking model receives 10 candidates retrieved by \sysemb and reorders them based on semantic relevance to the query (detailed in \S\ref{sec:data-synthesis}). Since the positive sample is guaranteed to be included in the candidates but not the first one (Rec@10 = 100\%, Rec@1=0\%), the reranking task focuses on promoting the correct binary code to higher positions with $k=\{1, 3\}$.

\begin{table}[htbp]
\vspace{-1.5ex}
\centering
\captionsetup{skip=2pt}
\small
\caption{The evaluation results of reranking models.}
\rowcolors{2}{gray!15}{white}
\setlength{\belowcaptionskip}{2pt}
\setlength{\tabcolsep}{2pt}
\begin{tabular}{@{}llccc@{}}
\toprule
Model                       & Sz.(B) & \multicolumn{1}{l}{$Rec@1$} & \multicolumn{1}{l}{$Rec@3$} & \multicolumn{1}{l}{$M\!R\!R@3$}   \\ \midrule
jina-reranker-v3            & 0.6      & 34.00                        & 63.75                        & 47.33                     \\
bge-reranker-v2-m3          & 0.6      & 34.00                        & 57.75                        & 46.03                     \\
Qwen3-Reranker-0.6B         & 0.6      & 35.00                        & 62.00                        & 46.42                     \\
Qwen3-Reranker-8B           & 8        & \textbf{62.75}               & \underline{81.00}            & \textbf{70.94}            \\
\sysrank                    & 0.6      & \underline{61.75}            & \textbf{83.00}               & \underline{70.23}         \\ \bottomrule
\end{tabular}
\label{tab:rerank-compare}
\vspace{-1.5ex}
\end{table}

As shown in \autoref{tab:rerank-compare}, our \sysrank achieved 61.75\% in Rec@1, meaning that the model successfully promoted the correct function to the top position in most cases. 
The Rec@3 score of 83,0\% indicates that users can locate the target function within the top-3 reranked results with high confidence. 
While the MRR@3 of 70.23\% further confirms that the correct results are typically ranked at the top positions after reranking.

Compared to general-purpose rerankers with the same parameter scale of 0.6B, \sysrank demonstrates substantial advantages. It outperforms Qwen3-Reranker-0.6B by 27.75\% of Rec@1 and 22.9\% of MRR@3. 
Moreover, \sysrank achieved competitive performance compared to Qwen3-Reranker-8B ($13\times$ larger), with 1\% and 0.71\% lower in Rec@1 and MRR@3 respectively, but 2\% leading in Rec@3. 
These results demonstrate that our domain training constructed effective and efficient expertise reranker.

Additionally, our context-aware design employed a heuristic informative measurement to select context functions as augmented information for improving reranking performance. Experiments demonstrate the effectiveness of this design, boosting our model and the baselines by 8\% and 4\% Rec@1, respectively. Details can be found in \S\ref{sec:ctx-comp}.

\vspace{-0.5ex}
\subsection{Overall Evaluation}
\vspace{-0.5ex}

To answer \textbf{RQ2}, we evaluated the overall performance of our framework \sysname, incorporating embedding and reranking models. In this evaluation, we performed reordering on all retrieved candidates without awareness of their labels and calculated the overall scores for the retrieval system. Additionally, we investigated variants composed of the same scale and the best-performing baseline models from previous evaluations.

As shown in \autoref{tab:pipeline-eval}, \sysname with our tailored models achieved the best performance, recording 76.75\% in Rec@1 and 84.5\% in Rec@3. Driven by embeddinggemma-300m and Qwen3-Reranker-0.6B that have the same model size as ours, the pipeline achieved 30.7\% lower in Rec@3 and 27.17\% lower in MRR@3, suggesting a significant effectiveness of our domain-specific training. 
Since the SFR-Embedding-Mistral and Qwen3-Reranker-8B have demonstrated the best performance among the evaluated embedding and reranking baselines, respectively, we thus employed them to investigate the best performance that can be expected from non-tailored models. 
As shown in the results, this combination achieved impressive scores, yet still lagged behind \sysname by 6.75\% and 4.63\% in Rec@3 and MRR@3, respectively. 
More importantly, considering the \texttt{20.93} minutes overhead for each query, which is 10 times longer than \sysname, it is challenging to be applied in practical scenarios. 
When working as a whole system, \sysname's Rec@1 score is 9.75\% higher than 67\% of the \sysemb standalone. This demonstrated that our 2-stage workflow design produces highly usable retrieval in practice. 


\vspace{-1ex}
\section{Conclusion}
\vspace{-1.0ex}
In this paper, we introduced \sysname, a two-stage cross-modal retrieval framework for the analysis of stripped binary code. 
It incorporated \sysemb to learn the semantic relevance between binary pseudocode and NL, and \sysrank to further refine the results with context augmentation.
For training our models, we also proposed an LLM-based data synthesis pipeline to generate large-scale high-quality data pairs automatically. 
Our evaluation demonstrated that \sysname achieved SOTA performance with 84.5\% in Rec@3 and 80.25\% in MRR@3, surpassing the best-performing general-purpose models by 6.75\% and 4.62\%, respectively, also demonstrating higher practical value with $10 \times$ faster speed.

\section{Limitations}

\noindent\textbf{Non-thorough Contextual Investigation.}
Although we incorporated callee functions to augment the semantic representation of binary code, our current approach is restricted to the local neighborhood of the call graph. Limited by the context length of the model, we heuristically selected top-$5$ callees as context augmentation. 
In practice, the semantics of a binary function could be intrinsically linked to its broader execution environment, including caller functions and deeper dependency chains. 
Thorough investigation of contextual information is essential for better performance in future work. 

\noindent\textbf{Limited Model Size.} 
Due to computational resource constraints, this study primarily validated the effectiveness of \sysname on lightweight architectures (0.3B for embedding and 0.6B for reranking). While these models achieved SOTA performance with superior efficiency, they may not fully leverage the emerging capabilities found in LLMs with billions of parameters. Future research will explore scaling up the model architecture to investigate the trade-offs between retrieval accuracy and computational overhead, as well as the potential of scaling laws in the binary analysis domain.

\section{Ethical Considerations}

While \sysname is designed to support defensive security workflows (e.g., vulnerability investigation, program auditing, and malware analysis by defenders), it may also lower the barrier to reverse engineering and be misused by malicious actors. In particular, by enabling natural language search over stripped binary codebases, our retrieval models could help attackers more quickly understand a target binary, locate security relevant functionality, and accelerate the analysis and development of malicious behaviors. Therefore, this technology can benefit legitimate security research and defense, but it also carries a risk of misuse.



\bibliography{main}

\appendix

\clearpage

\section{Appendix}
\label{sec:appendix}

\subsection{Raw Dataset Details}
\label{sec:appendix-raw-data}
We selected C/C++ open-source projects based on three key criteria: active maintenance, popularity within the developer community, and domain diversity.
Specifically, within the GitHub community, we prioritized projects exhibiting recent commit activity and high star counts, ensuring broad coverage across fields such as cryptography, networking, multimedia, and databases.

To build diverse binaries, we utilized the \texttt{makepkg}~\cite{makepkg} pipeline equipped with wrapped compilers. \texttt{makepkg} is a script to automate the building of packages released by Arch Linux. We wrapped the \texttt{gcc} and \texttt{clang} compilers to customize the compilation process with custom build flags, such as optimization levels and debug information. 
Although the compilation process is automated by scripts, it is still prone to errors, such as dependency issues or configuration mismatches.
Therefore, we further integrated AutoCompiler~\cite{ACL2025compileagent}, which is an LLM agent-based automated compilation pipeline, to resolve build failures and expand our collection. 
We stored the compiled debugging information into a distinct file, which is used to link a stripped binary function to its source code.

We employed configurations covering optimization levels ranging from \texttt{O0} to \texttt{O3}, utilizing \texttt{gcc-11} and \texttt{clang-13} as compilers. We targeted a wide array of CPU architectures, including \texttt{x86-32}, \texttt{x86-64}, \texttt{arm32}, \texttt{arm64}, \texttt{mips32}, and \texttt{mips64}.
Ultimately, we successfully compiled \texttt{10,555} open-source projects into artifacts, yielding \texttt{1,817,461} binary files. These binaries were decompiled into \texttt{183,925,317} pseudocode functions, with approximately 58\% successfully aligned to their corresponding source code.

\subsection{Data Synthesis Details}
\label{sec:appendix-data-syn}

As mentioned in \S\ref{sec:data-synthesis}, we employ an advanced large language model to generate semantic descriptions and assess their quality. Specifically, we use DeepSeek-V3~\cite{2025deepseekv3}, a SOTA open-source LLM with 671B parameters, to perform both generation and quality evaluation tasks. Here we present the details of these prompts:

\noindent\textbf{Generation Prompt.}
As shown in \autoref{fig:generation-prompt}, the prompt provides the LLM with the complete source code file content along with contextual information including the file path, project name, and version number. This contextual information helps the model better understand the function's role within the project structure. We explicitly require the model to generate concise one-sentence descriptions that capture the core functionality of each function. Additionally, we ask the model to provide both English and Chinese versions of the description in JSON format, which facilitates subsequent processing and enables bilingual support for our dataset.

\begin{figure}[t]
\centering
\begin{minted}[
    breaklines,
    fontsize=\small,
    frame=lines,
    bgcolor=lightgray!10
]{markdown}
# Role
Imagine you are an experienced software developer.

# Task
The user will provide a source code function and its basic information each time.
Your task is to analyze its semantics and generate a comment to the function.
The comment should be a brief description of the function in one sentence.

# Rules
When you are generating the comment, please follow the rules below:

1. Comment should be accurate, precise, and helpful for code understanding.
2. You can refer to the original comments in the source code, but you cannot directly copy the original comments.
3. You also need to provide a Chinese version, which should have the same meaning as the English version.
4. You need to write comments into a JSON format string, for example:
```json
{
    "en": "Brief description of the function in one sentence",
    "cn": "The chinese version of the description",
}
```

# Input

Here is a source code from {path} file in the {project} {version} project:
```C/C++
{code}
```

You should generate a comment for the function named `{func_name}`.

Start generating
\end{minted}
\caption{The prompt for generating semantic descriptions based on source code functions.}
\label{fig:generation-prompt}
\end{figure}

\noindent\textbf{Discrimination Prompt.}
\autoref{fig:discrimination-prompt} presents the discrimination prompt, which instructs the model to assess the quality of generated descriptions based on three criteria: accuracy, relevance, and conciseness. We establish a four-level rating scale (A to D) with clear definitions: \textit{Excellent (A)} for perfectly accurate descriptions, \textit{Good (B)} for generally accurate ones with minor issues, \textit{Fair (C)} for vague or slightly inaccurate descriptions, and \textit{Poor (D)} for factually incorrect or irrelevant content. The model is required to output both a score and a brief explanation in JSON format. In our Generation Filter described in \S\ref{sec:data-synthesis}, we only retain data pairs with ratings of A or B, effectively filtering out low-quality descriptions that may contain errors, hallucinations, or irrelevant information.

\begin{figure}[t]
\centering
\begin{minted}[
    breaklines,
    fontsize=\small,
    frame=lines,
    bgcolor=lightgray!10
]{markdown}
# Role
Imagine you are an experienced software developer and code reviewer.

# Task
The user will provide a C/C++ source code function and a generated natural language description (comment) for it.
Your task is to evaluate the quality of the generated description based on its accuracy, relevance, and conciseness.

# Rules
Please rate the description on a scale from A to D based on the following criteria:

- **A (Excellent)**: The description is accurate, precise, and captures the core functionality perfectly without errors.
- **B (Good)**: The description is generally accurate and helpful, but may miss minor details or contains slight redundancy.
- **C (Fair)**: The description is vague, too generic, or contains minor inaccuracies/irrelevant content.
- **D (Poor)**: The description is factually incorrect, completely irrelevant, hallucinations, or consists of meaningless repetition.

You need to write the evaluation result into a JSON format string:
```json
{
    "score": "A", // Options: A, B, C, D
    "reason": "Brief explanation of the rating"
}
```

# Input

Here is a source code from {path} file in the {project} {version} project:
```C/C++
{code}
```

The comment is generated for the function named `{func_name}`.

Here is the generated description:
```
{comment}
```

Start evaluating
\end{minted}
\caption{The prompt for evaluating the quality of generated function semantic descriptions.}
\label{fig:discrimination-prompt}
\end{figure}

\subsection{Human Evaluation on Synthetic Labels}
\label{sec:human-eval}

To validate the quality of the LLM-generated natural language descriptions used for training, we conducted a human evaluation on a randomly sampled subset of the synthesized dataset.
We randomly selected 100 function–description pairs. For each sample, human evaluators were provided with the original source-level function along with the same meta information that was available to the LLM during label generation, while the LLM discriminator's assessment was not shown.

Three human evaluators independently participated in the evaluation: one reverse engineering expert with eight years of industry experience, and two graduate students with four and three years of experience in binary analysis, respectively.
To mitigate potential fatigue and mechanical judgments, each evaluator assessed 20 samples per day, spending about 30 minutes per session.

Evaluators were asked to judge ``\textit{whether the generated description accurately reflects the core functionality of the corresponding function, using three categories: Correct, Partially correct, or Incorrect}''.
Across all evaluators, an average of 90.3\% of the generated descriptions were rated as Correct, and the percentage increases to 97.6\% when including Partially correct cases. This indicates that the LLM-generated labels are largely accurate and provide reliable supervision for model training.

\subsection{Dataset Construction Algorithms.}
\label{sec:appendix-data-const-algo}
We present the detailed procedures for constructing our evaluation benchmarks in Algorithm~\ref{alg:embedding_eval} and Algorithm~\ref{alg:reranking_eval}.

Algorithm~\ref{alg:embedding_eval} outlines the construction of the embedding evaluation dataset. It employs an iterative sampling strategy with a similarity-based filter (\texttt{MaxSim}) to collect real negative samples that are semantically distinct from the query, ensuring a robust assessment of retrieval capability.

Algorithm~\ref{alg:reranking_eval} details the creation of the reranking evaluation dataset. This process involves retrieving candidates using the trained embedding model and selecting instances where the positive sample is retrieved but not ranked first, thereby simulating realistic scenarios for evaluating the reranker's refinement ability.


\begin{algorithm}[t]
\caption{Embedding Evaluation Dataset Construction.}
\label{alg:embedding_eval}
\begin{algorithmic}[1]
\Require Corpus $\mathcal{C}$; Pool size $K$; Threshold $\rho_{\text{th}}$
\Ensure Benchmark $\mathcal{B} = \{(q_i, \mathcal{P}_i)\}_{i=1}^{N}$
\State $\mathcal{B} \leftarrow \emptyset$
\For{each $(q_i, p_i) \in \mathcal{C}$}
    \State $\mathcal{P}_i \leftarrow \{p_i\}$
    \While{$|\mathcal{P}_i| < K$}
        \For{$p_j \in \mathcal{C} \setminus \mathcal{P}_i$}
            \State $\rho \leftarrow \texttt{MaxSim}(q_i, q_j)$
            \If{$\rho < \rho_{\text{th}}$}
                \State $\mathcal{P}_i \leftarrow \mathcal{P}_i \cup \{p_j\}$
                \If{$|\mathcal{P}_i| = K$}
                    \State \textbf{break}
                \EndIf
            \EndIf
        \EndFor
    \EndWhile
    \State $\mathcal{B} \leftarrow \mathcal{B} \cup \{(q_i, \mathcal{P}_i)\}$
\EndFor
\State \Return $\mathcal{B}$
\end{algorithmic}
\end{algorithm}

%

\begin{algorithm}[htbp]
\caption{Reranking Evaluation Dataset Construction.}
\label{alg:reranking_eval}
\begin{algorithmic}[1]
\Require Embedding eval. data $\mathcal{B}$; Embedding model $M_e$; Retrieval size $N$
\Ensure Reranking benchmark $\mathcal{R} = \{(q_i, \mathcal{R}_i)\}$
\State $\mathcal{R} \leftarrow \emptyset$
\For{each $(q_i, \mathcal{P}_i) \in \mathcal{B}$}
    \State $\mathcal{C}_i \leftarrow M_e\text{.retrieve}(q_i, \mathcal{P}_i, N)$
    \If{$p_i \in \mathcal{C}_i$ and $p_i \neq \mathcal{C}_i[0]$}
        \State $\mathcal{N}_i \leftarrow \text{TopRanked}(\mathcal{C}_i \setminus \{p_i\}, N-1)$
        \State $\mathcal{R}_i \leftarrow \{p_i\} \cup \mathcal{N}_i$
        \State $\mathcal{R} \leftarrow \mathcal{R} \cup \{(q_i, \mathcal{R}_i)\}$
    \EndIf
\EndFor
\State \Return $\mathcal{R}$
\end{algorithmic}
\end{algorithm}





\begin{table*}[t]
\centering
\caption{Details of our \sysname and the selected baselines for embedding-based retrieval and reranking.}
\setlength{\tabcolsep}{2pt}
\begin{tabular}{@{}lllll@{}}
\toprule
Model                          & Domain & Model Size & Emb. Dim. & Max Token \\ \midrule
                               & \multicolumn{4}{c}{Retrieval Model}         \\ \midrule
embeddinggemma-300m            & Text   & 0.3B       & 768       & 2048     \\
Qwen3-Embedding-0.6B           & Text   & 0.6B       & 1024      & 32768     \\
Qwen3-Embedding-8B             & Text   & 8B         & 4096      & 32768     \\
multilingual-e5-large-instruct & Text   & 0.6B       & 1024      & 512       \\
SFR-Embedding-Mistral          & Text   & 7B         & 4096      & 4096      \\
BGE-M3                         & Text   & 0.6B       & 1024      & 8192      \\
SFR-Embedding-Code-400M        & Code   & 0.4B       & 1024      & 8192      \\
jina-embeddings-v4             & Code   & 3B         & 2048      & 32768     \\
inf-retriever-v1               & Code   & 7B         & 3584      & 32768     \\
\sysemb                        & Binary & 0.3B       & 1024      & 4096      \\ \midrule
                               & \multicolumn{4}{c}{Reranking Model}         \\ \midrule
Qwen3-Reranker-0.6B            & Text   & 0.6B       & /         & 32768     \\ 
Qwen3-Reranker-8B              & Text   & 8B         & /         & 32768     \\ 
BGE-Reranker-v2-m3             & Text   & 0.6B       & /         & 8192      \\
jina-reranker-v3               & Text   & 0.6B       & /         & 131072    \\
\sysrank                       & Binary & 0.6B       & /         & 16384     \\ \bottomrule
\end{tabular}
\label{tab:baselines}
\end{table*}

\subsection{Baseline Model Specifications.}
\label{sec:appendix-model-select}

As mentioned in \S\ref{sec:eval-settings}, we carefully selected 13 baseline models for comprehensive evaluation: 9 embedding models and 4 reranking models. Since there is no prior work directly addressing NL-pseudocode retrieval for stripped binary analysis, we compare \sysname against general-purpose models that represent the current SOTA. Our selection strategy focuses on three key criteria: (1) top performance on public benchmarks, (2) popularity within the community, and (3) accessibility for reproduction. 

For embedding models, we primarily draw from two well-known leaderboards: MMTEB~\cite{2025mmteb} and CoIR~\cite{CoIR} (records as of December 2025). From MMTEB, we selected top-ranked general-purpose text embedding models including Qwen3-Embedding~\cite{2025qwen3-emb}, Jina-Embeddings~\cite{2025jinaembeddingsv4}, multilingual-e5-large~\cite{wang2024multilingual}, and embeddinggemma~\cite{vera2025embeddinggemma}. From CoIR, which specifically evaluates code-related retrieval tasks, we selected specialized models including SFR-Embedding~\cite{liu2024codexembed,2024SFR}, BGE-M3~\cite{2024bge-m3}. 
As for reranking models, we selected four top-performing models from the MMTEB leaderboard~\cite{2025mmteb}: Qwen3-Reranker~\cite{2025qwen3-emb}, BGE-Reranker-v2-m3~\cite{2024bge-m3}, and Jina-Reranker-v3~\cite{wang2025jinarerankerv3}. \autoref{tab:baselines} lists the detailed specifications of all baseline models, including parameter size, embedding dimension, and maximum context length, facilitating a fair and transparent comparison.

\subsection{Ablation Studies on Binary Code Context}
\label{sec:ctx-comp}

\begin{figure}[htbp]
    \centering
    \includegraphics[width=1\linewidth]{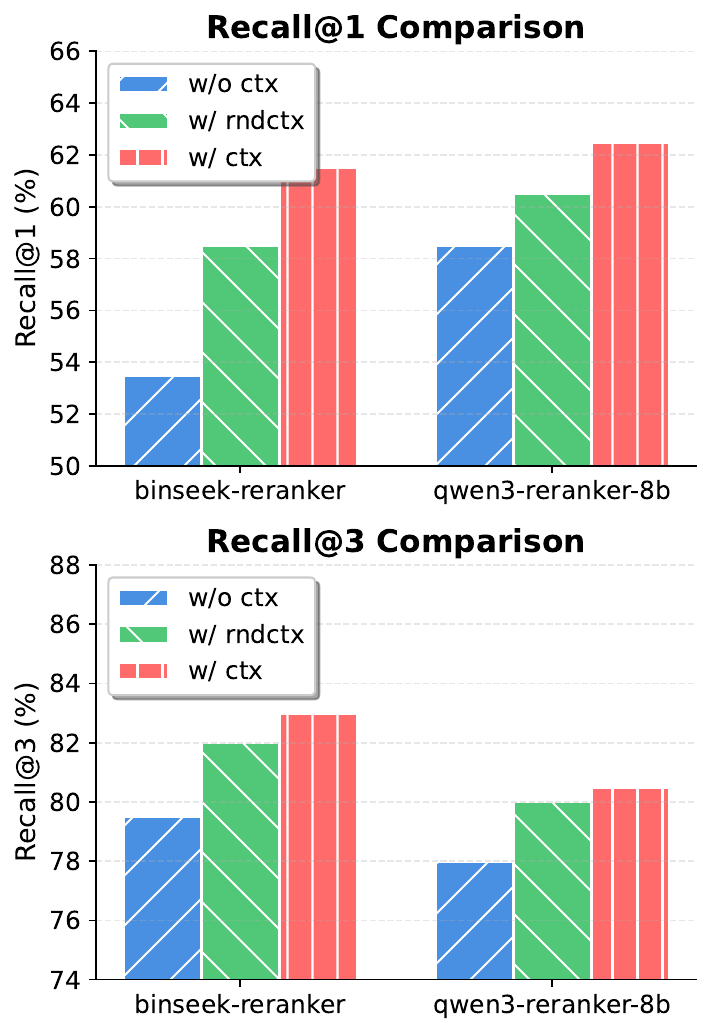}
    \caption{Comparison of the selection method of binary code context across different metrics and rerankers. ``\texttt{ctx}'' means the context functions are selected by our heuristic informative score $\mathcal{S}$, and ``\texttt{rndctx}'' means randomly selecting context functions. While ``\texttt{w/o ctx}'' means use no context in reranking.}
    \label{fig:ctx-comp}
\end{figure}

To further investigate the effectiveness of context augmentation and our heuristic informative measurement, we conducted an ablation study on the context selection method. We compared the performance of the following three methods: (1) using our heuristic informative measurement to select the context functions with top-$5$ scored, (2) randomly selecting $5$ context functions, and (3) using no context in reranking. We involved our \sysrank and a representative baseline model Qwen3-Reranker-8B in this experiment.

As shown in \autoref{fig:ctx-comp}, both \sysrank and Qwen3-Reranker-8B achieved the best performance when using our heuristic informative measurement to select the context functions. In particular, the selected context boosted our model and the baseline model with 8\% and 4\% Rec@1, respectively. Meanwhile, randomly selecting context functions also achieved better performance than using no context. This demonstrates the effectiveness of our context-aware design. On the other hand, randomly selected context achieved suboptimal performance compared to employing our selection method, suggesting the contribution of our informative measurement. 

\end{document}